\documentclass[10pt,aps,prd,superscriptaddress,twocolumn,amsmath,amssymb,floatfix,
nofootinbib 
]{revtex4-2} 

\usepackage[dvips]{graphics}
\usepackage{bm}
\usepackage{epsfig}
\usepackage{enumerate}
\usepackage{subfigure}
\usepackage{color} 
\usepackage{graphicx}
\usepackage[dvipsnames]{xcolor}
\usepackage{graphicx}
\usepackage[colorlinks,citecolor=blue,linkcolor=blue,urlcolor=blue,plainpages=false,pdfpagelabels]{hyperref}
\usepackage{orcidlink}
\usepackage{comment}
\newcommand{\beq}{\begin{equation}}
\newcommand{\eeq}{\end{equation}}
\newcommand{\bea}{\begin{eqnarray}}
\newcommand{\eea}{\end{eqnarray}}

\begin{document} 
\title{Persistent nonlinear Hall effect driven by parallel field across a topological phase transition and intraband sign-reversing integer quantum Hall effect} 

\author{Suheel Ahmad Malik \orcidlink{0009-0000-1618-6602 }}
\email{malik.suhail08@gmail.com}
\affiliation{Department of Physics, Jamia Millia Islamia, New Delhi-110025, INDIA}
\author{M.A.H. Ahsan \orcidlink{0000-0002-9870-2769}}
\affiliation{Department of Physics, Jamia Millia Islamia, New Delhi-110025, INDIA}
\author{SK Firoz Islam \orcidlink{0000-0003-1224-622X}}
\email{ s_islam2@jmi.ac.in}
\affiliation{Department of Physics, Jamia Millia Islamia, New Delhi-110025, INDIA}
\date{\today}

\begin{abstract}
We investigate the linear and nonlinear Hall effects in a two-dimensional Rashba spin-orbit coupled nodal ring electronic system. We consider both the cases of linear Hall effect, the Berry curvature induced anomalous Hall and the perpendicular magnetic field-induced integer quantum Hall effect. We observe that the system exhibits a topological gap along the boundary of the nodal ring that strongly depends on the radius of the nodal ring and the strength of the Rashba spin-orbit interaction, resulting in the quantum anomalous Hall effect. Subsequently, we include a perpendicular uniform magnetic field and obtain the exact Landau levels that exhibit sign reversal in the slope with magnetic field around the ring boundary. This causes a sign reversal in the quantum Hall conductivity in the same band by tuning the magnetic field. Most importantly, we also show that when the magnetic field is strictly parallel to the system, it can induce an anisotropy to the Berry curvature that leads to the emergence of the nonlinear Hall effect. Additionally, tuning the parallel field can close and reopen the topological gap, resulting in the reverse topological phase transition from the topological Chern insulator to a trivial insulator. The nonlinear Hall effect remains persistent in both phases, but its dependence on the chemical potential exhibits distinct signatures. Noticeably, the nonlinear Hall response exhibits a sign-changing peak structure in Chern insulating phase whereas it displays a single-sign response in trivial insulating phase. These distinct signatures suggest that nonlinear Hall can serve as a probe to study such a topological phase transition.
\end{abstract}  
\maketitle
\section{ Introduction} In recent years, the concept of quantum geometry, comprising the Berry phase and quantum metric, has emerged as an important framework for understanding both linear and nonlinear transport phenomena in condensed matter systems \cite{RevModPhys821959,Torma2023,Yu2025}. The Berry curvature acts as an effective magnetic field and gives rise to an anomalous Hall response in systems with broken time-reversal symmetry (TRS), even in the absence of an external magnetic field \cite{Haldane2004,Xiao2010,Nagaosa}. However, the phenomenon of the quantum Hall effect traces back to the discovery of the integer quantum Hall effect in the presence of a perpendicular magnetic field by von Klitzing et al. in 1980 \cite{Klitzing1980}. The core underlying physics in both the cases is the breaking of TRS \cite{Haldane2004}. It is noteworthy that the Berry curvature resulting from the breaking of inversion symmetry (IS) does not give rise to the anomalous Hall response. However, recently, Sodemann and Fu proposed that the first-order moment of the Berry curvature, viz., the Berry curvature dipole (BCD), can give rise to a nonlinear Hall response even in time-reversal invariant materials lacking IS \cite{sodemann2015,Ortix2021,Du2021,Bandyopadhyay2024}. Of course, breaking IS is not sufficient to achieve the nonlinear Hall, an additional symmetry, like rotational or mirror needs to be broken. Following the theoretical proposal, BCD-induced nonlinear Hall effect was experimentally observed in bilayer and few layer WTe$_2$\cite{Ma2019,Kang2019}. Motivated by these experimental observations, BCD-induced nonlinear Hall effect was further theoretically investigated in various materials, which include transition metal chalcogenides\cite{You2018,Joseph2021}, Weyl semimetals\cite{Zhang2018,Chuanchang2021}, etc. In pristine materials, the crystall symmetries that prohibit finite BCD-induced nonlinear Hall can also be broken externally through various perturbative mechanisms. Therefore, various external mechanisms, including strain \cite{Araki2018,PhysRevLett.123.196403}, Twist \cite{Chakraborty2022-bw}, periodic drive \cite{Chen2022,Qin2024,zhu2026}, electric field \cite{Xu2018,Ye2023} etc. have also been explored to overcome these symmetries and cause finite BCD-induced nonlinear Hall response.


\par The Rashba spin-orbit coupling (RSOC) is a relativistic effect \cite{Winkler2003} that not only splits up the energy band into two spin branches but also causes spin-momentum locking. The RSOC is the key ingredient of spin-Hall effect \cite{RevModPhys.87.1213}. The RSOC arises from the lack of IS, making such a system a suitable platform for realizing the nonlinear Hall effect. In a usual two-dimensional electron gas formed across the semiconductor heterojunction, the RSOC band structure does not possess any Berry curvature and band anisotropy, no nonlinear Hall effect is expected. Hence, such a system requires some additional symmetry-breaking perturbations. Recently, the hexagonal warping term has been considered in Ref.~[\onlinecite{Saha2023}] to achieve the nonlinear Hall. Beyond the conventional Rashba systems, the nonlinear Hall response has recently been studied in unconventional Rashba systems, where electric field-induced Berry curvature polarizability gives rise to a finite nonlinear Hall response \cite{ankita2025}. It is noteworthy to mention here that in a spin-momentum locking electronic system, parallel magnetic field has also been known to play crucial role in enhancement of spin-Hall effect \cite{PhysRevB.71.085315}, quantum phase transition \cite{PhysRevB.83.245428}, planer Hall effect \cite{Taskin2017} etc.


\par Recently, few theoretical predictions have been made on the existence of $2$D and $3$D magnetic nodal ring semimetals with RSOC that is linear in momenta\cite{josephson2025,74kyd71n}. The exciting aspect of such a system is that the spin transport and band topology are strongly sensitive to the radius of the nodal ring. In this work, we consider such a system with RSOC term that has recently been considered by several studies \cite{PhysRevLett.123.116401,PhysRevB.102.125118,PhysRevB.99.035125,josephson2025}. In fact, even without the RSOC term, the $2$D nodal ring semimetal has been considered in a spinless system in studying the magnetotransport properties.  \cite{PhysRevB.111.075125}. 

First, we analyze the system's topology and note that it exhibits a topological gap around the boundary of the nodal ring, making it a Chern insulator. As expected, the system exhibits linear anomalous quantum Hall effect. Further, we consider another route to realizing the linear quantum Hall effect by applying a perpendicular magnetic field — the integer quantum Hall effect. We observe a peculiar kind of spin-resolved Landau levels (LLs), exhibiting sign reversal in the slope with magnetic field across the ring boundary. This unusual nature of LLs leads to a sign reversal in the quantum Hall conductivity. Finally, and most importantly, we show that applying a parallel magnetic field can give rise to an anisotropic Berry curvature, resulting in the BCD-induced nonlinear quantum Hall effect. Not only that, with the tuning of the parallel field, the Chern insulating phase undergoes a trivial insulating phase by closing and reopening the gap. Interestingly, the nonlinear Hall effect remains persistent across both sides of the topological phase transition as long as the anisotropic Berry curvature remains. We also find that the nonlinear Hall response exhibits distinct behavior as a function of the chemical potential in the two phases. Therefore, the nonlinear Hall effect can itself serve as an effective probe for identifying such topological phase transitions.

The manuscript is presented in four sections. The Sec.~(\ref{band_structure}) introduces the system's Hamiltonian and corresponding spin-resolved band structure. The quantum-Hall effect, comprising the linear as well as the nonlinear Hall effect, is presented in Sec.~(\ref{Quantum_hall}). Finally, we summarise in Sec.~(\ref{summary}). 

\section{ Model Hamiltonian and band structure}\label{band_structure}
We start with the low-energy effective Hamiltonian for the system as \cite{josephson2025} $H=  {\bf d}\cdot{\bf\sigma}$ where, $\{d_x,d_y,d_z\}=\{\alpha k_y,-\alpha k_x,\epsilon( k^2 - R^2)\} $ . Here, $\epsilon$ is an inverse mass term, ${\bf k}=\{k_x,k_y\}$ is $2$D momentum operator, $R$ is the radius of the nodal ring and ${\bf \sigma} = \{\sigma_x,\sigma_y,\sigma_z\}$ are Pauli matrices in real spin basis. The terms $\{d_x,d_y\}$ correspond to RSOC, with strength $\alpha$. Note that this RSOC term is the usual linear in $k$ type as exists in a conventional $2$D electronic system \cite{Winkler2003}. The energy spectrum can be obtained by diagonalizing the above Hamiltonian as $E_{{\bf k},\lambda}= \lambda|\bf d|$ where $\lambda=\pm$ is the band index. The corresponding wave function is given by
\begin{equation}
\Psi_{k,\lambda}({\bf r})=\frac{\exp({i\bf{k.r}})}{\sqrt{A}}\left[\begin{array}{c}1 \\ \frac{d_x+id_y}{\lambda|d|+d_z}\end{array}\right]
\end{equation}
with $A$ being the area of the system. It is worth noting that, unlike the conventional $2$D electron gas with RSOC, in this system spin does not lie entirely in the $xy$-plane, it has an out-of-plane component, as indicated by the $\sigma_z$ term in the Hamiltonian.

\begin{figure}
{\includegraphics[width=0.99\linewidth]{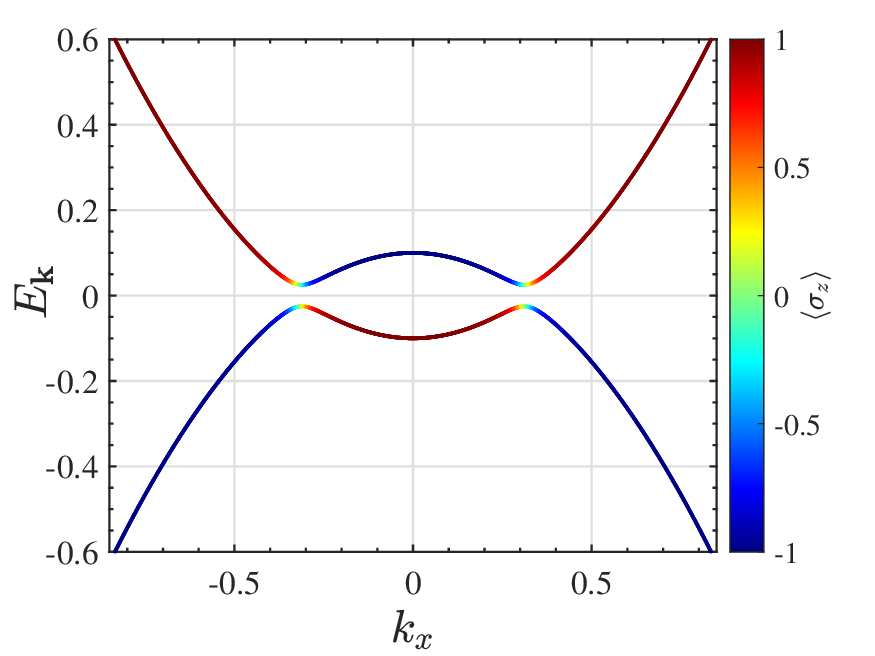}}
\caption{Energy spectrum of the nodal ring semimetal with RSOC. The momentum axis is normalized by a typical Fermi wave vector $k_F$ corresponding to standard $2$D electron density, whereas the energy axis is normalized by $\epsilon k_F^2$. The band gap is shown along the radius at $k=R$. The magnitude of out-of-plane spin polarization is shown by the color bar. Parameters taken are $R=0.3$ and $\alpha=0.08$ in the unit of $k_F$ and $\epsilon k_F$}
\label{spectrum1}
\end{figure}

The out-of-plane spin polarization in momentum space is given by the $z$-component of average spin polarization that can be obtained as  $\langle\sigma_z\rangle\propto\lambda\epsilon \operatorname{sgn}(k^2-R^2)$ is sensitive to the band index. Similarly, the in-plane spin polarization can also be written as $\langle\sigma_{x}\rangle\propto \operatorname{sgn}(k_y)$ and $\langle\sigma_y\rangle\propto \operatorname{sgn}(k_x)$, which does not depend on the band index. Using the out-of-plane spin polarization component, we plot spin-resolved energy spectrum in Fig.~\ref{spectrum1}, which shows a band gap along the nodal ring boundary given by $\Delta=2\alpha R$. Also, the two bands have opposite out-of-plane spin polarization both inside and outside the nodal ring. However, in the same band, the $\langle\sigma_z\rangle$ gets flipped across the radius of the ring without affecting the in-plane spin components.

\section{Quantum Hall effect}\label{Quantum_hall}

The quantum Hall effect generally refers to the quantization of the Hall conductivity in a TRS-broken $2$D fermionic system. However, quantization of the Hall response is restricted to the linear quantum Hall effect, the nonlinear quantum Hall effect does not show any quantization. In order to understand the linear and nonlinear Hall effects, let us briefly discuss the electric current density in the system in the presence of an applied in-plane electric bias. The current density $J_{a}$ for a $2$D electronic system along the $a$ direction can be written as
\begin{equation}  \label{eqn:current_den}
J_{a}= -e\int \frac{d^2k}{(2\pi)^2} f_k~v_{a}
\end{equation}
where, $f_{k} \equiv f({\bf r},k,t)$ is the non equilibrium distribution function and $v_{a}$ is the velocity of the electrons along the $a$ direction. In the presence of the external electric field, the velocity of the electron consists of two parts, the normal band velocity that arises from the semiclassical wave packet and the anomalous velocity contribution arising from the Berry curvature. Therefore, we can write  
\begin{equation}\label{velocity}
v_a =\frac{1}{\hbar} \frac{\partial E_{k}}{\partial k_a} + \epsilon_{abc} \dot{k}_{b}\Omega_{c} 
\end{equation}
  where the time evolution of the crystal momenta $\dot{\bf k}=-e {\bf \mathcal{ E}}/\hbar$ where ${\bf\mathcal{E}}$ is the applied bias. Also, $E_{k}$ denotes the energy with momenta $k$, and $\Omega_c$ denotes the $c$-component of the Berry curvature \cite{RevModPhys821959}. Also, in the presence of the external perturbation, the distribution function follows the semiclassical Boltzmann equation. In the relaxation time approximation, the semiclassical Boltzmann transport equation for a spatially uniform system under the steady state can be written as $ f_{k}=
f^{(0)}_{k} + (e\tau/\hbar)  \boldsymbol{\mathcal{E}}\cdot \nabla_{\bf k} f_{k}$, where  $ f^{(0)}_{k}= [\exp\{(E_{k}-\mu)/k_{B}T\}+1]^{-1} $ is the equilibrium distribution function with $T$ as temperature, $k_B$ as the Boltzmann constant and $\mu$ as the chemical potential. Here, $\tau$ is the relaxation time. It can be solved perturbatively by expanding in the powers of the electric field as $f_{k} = f_{k}^{(0)} + f_{k}^{(1)} + f_{k}^{(2)} + \cdots$. Since we are interested in the second-order response, expanding $f_{k}$ upto linear order, we obtain

\begin{equation} \label{eqn:fermi_exp}
f_{k}= f_{k}^{(0)}+\frac{e\tau}{\hbar} \boldsymbol{\mathcal{E}}\cdot \nabla_{\bf k}f_{k}^{(0)}.
\end{equation}
Now plugging Eq.~(\ref{eqn:fermi_exp}) and (\ref{velocity}) into Eq.~(\ref{eqn:current_den}), we obtain the current density up to the second order in $\mathcal{E}$ within the relaxation time approximation and  under the steady state condition as
\begin{equation}
J_{a}= \sigma_{ab} \mathcal{E}_{b} + \chi_{abc}\mathcal{E}_{b} \mathcal{E}_{c}.
\end{equation}
The coefficient $ \sigma_{ab} $ in the first term describes the linear Hall response, while the second term coefficient $\chi_{abc}$ describes the second-order anomalous Hall response. For a multiband system, the second-order coefficient is obtained by summing over the band index $n$ and is given as \cite{sodemann2015}
\begin{equation}
\chi_{abc}=\epsilon_{acd} \frac{e^3\tau}{2\hbar^2} \sum_{n} \int \frac{d^2k}{(2\pi)^2} v_{b}^\zeta\Omega_{c}^\zeta~\frac{\partial{f_{\zeta}^{(0)}}}{\partial{E_{\zeta}}}
\end{equation}
where, $\epsilon_{acd}$ is the Levi-Civita tensor, $(a,b,c,d) \in (x,y,z) $ the integral part of this equation is known as BCD and is defined as
\begin{equation}   \label{eqn:bcd}
\mathcal{D}_{bd}=\sum_{n} \int \frac{d^2k}{(2\pi)^2} v_{b}^\zeta\Omega_{d}^\zeta~\frac{\partial{f_{\zeta}^{0}}}{\partial{E_{\zeta}}}.
\end{equation}
Here, $\zeta\equiv\{n,k\}$ denotes the quantum number. For a system with broken TRS the above momentum integration vanishes. However, it survives for an IS symmetry broken system with a band anisotropy. 

\subsection{Linear quantum Hall effect}
The linear Hall effect originates from the TRS breaking gap induced Berry curvature. If the system has intrinsically broken TRS without any external magnetic field, then the Hall response is known as the intrinsic anomalous Hall effect. However, if an external magnetic field perpendicular to the $2$D system is applied to break the TRS, then the corresponding Hall phenomenon is known as the integer quantum Hall effect. We present both cases individually.

\subsubsection{Intrinsic anomalous Hall effect}
Here, we present the quantum anomalous Hall response that arises in the absence of a real magnetic field. We briefly perform symmetry analysis as well as the topological nature of the band gap of the Hamiltonian. The Hamiltonian of the system $H$, as introduced earlier, is not symmetric under time reversal i.e., $\mathcal{T}H\mathcal{T}^{-1}\neq H$, as under time reversal ${\bf \sigma}\rightarrow-{\bf \sigma}$, ${\bf k}\rightarrow -{\bf k}$. Hence, the gap term $\Delta=2\alpha R$ emerges along the boundary of the nodal ring, which is topological in nature, and can be further confirmed by looking into the topological invariant-Chern number $\mathcal{C}=\int_k\Omega(\bf{k})$ where $\Omega({\bf k})$ is the topological gap induced Berry curvature and the momentum integration is taken over the filled band. Using the standard formula of Berry curvature for a two-band system as \cite{RevModPhys821959} \begin{equation}
\Omega({\bf k}) = -\lambda\frac{1}{2|\bf d|^3} {\bf d} \cdot\bigg(\frac{\partial{\bf d}}{\partial k_x}\times\frac{\partial{\bf d}}{\partial k_y}\bigg),
\end{equation}
we get $\Omega({\bf k})=\alpha^2\epsilon(k^2+R^2)/(2|{E}_{k,\lambda}|^3)$ leading to the non-zero Chern number $\mathcal{C}= 1$ indicating the non-trivial topological phase of the system that belongs to the class of topological Chern insulator. Here, we set the $\lambda=-1$ as the chemical potential fully occupies the valence band only. The anomalous quantum Hall conductivity can be immediately written as $\sigma_{xy}^{\rm AH}= e^2/h$ when the chemical potential lies inside the gap and occupies the valence band completely.
 
 \subsubsection{Integer quantum Hall effect}
 Now we consider another case of the linear quantum Hall effect--integer quantum Hall effect. Research investigation on the integer quantum Hall effect continues to thrive in advanced materials such as graphene\cite{GusyninPRL}, $8$-Pmmn borophene\cite{Islam2018}, the $\alpha$-$T_3$ lattice\cite{Tutul2016}, and most recently in unconventional RSOC systems\cite{Aryan2026}, etc. Note that the present system is already broken under time reversal, hence one can expect that the application of an additional TRS breaking perturbation like magnetic field can increase the number of Hall edge modes by forming LLs. {

Let's consider that the constant perpendicular magnetic field ${\bf B}=B\hat{z}$ is applied to the system lying in $xy$-plane. The field is included in the Hamiltonian via the Peierls substitution ${\bf k}\rightarrow {{\bf k}+e{\bf A}/\hbar}$ where ${\bf A}$ is the magnetic vector potential. Using  Landau gauge $\mathbf{A}=(0,xB,0)$, which yields
\begin{align}     \label{eqn:H2}
H=\epsilon\left[k_x^2+\left(k_y+\frac{eBx}{\hbar}\right)^2-R^2\right]\sigma_z \nonumber\\+ \alpha\left[\left(k_y+\frac{e Bx}{\hbar}\right)\sigma_x-k_x\sigma_y\right].
\end{align}
The Hamiltonian is translationally invariant along the $y$-direction as $[H,k_y]=0$, therefore the total wave function can be written as $\Psi(x,y)= e^{i k_yy}\phi(x)/\sqrt{L_y}$, with $L_y$ as the size of the system along the $ y$-direction. We introduce the ladder operators as $\hat{a}=(\tilde{x}+\partial/\partial\tilde{x})/\sqrt{2}$ and $a^\dagger=(\tilde{x}-\partial/\partial\tilde{x})/\sqrt{2}$. The position operator is defined as $\tilde{x}=(x+x_{0})/l_{c}$ with centre of cyclotron orbit $x= {-x_{0}}=- l_{c}^2k_{y}$ where, $l_{c}=\sqrt{\hbar/eB}$ is the magnetic length and the dimensionless $x$-component of the momentum operator $\tilde{k}_x=-i\partial /\partial(x/l_c)$. Using this, the above Hamiltonian takes the form
\begin{equation}
H=
\begin{bmatrix}
\hbar\omega_c\left(\hat{a}^\dagger\hat{a}+\frac{1}{2}\right)-\epsilon R^2&\beta \hat{a}\\
\beta \hat{a}^\dagger& -\hbar\omega_c\left(\hat{a}^\dagger\hat{a}+\frac{1}{2}\right)+\epsilon R^2
\end{bmatrix},
\end{equation}
where $\omega_c=2e\epsilon B/\hbar^2$ is the cyclotron frequency and $\beta=\sqrt{2}\alpha/l_c$ is the RSOC energy scale. We solve the secular equation $H\phi(x)=E\phi(x)$ to obtain the energy eigenvalues and corresponding eigenfunction.
\begin{figure}[t]
\centering
\includegraphics[width=0.90\linewidth]{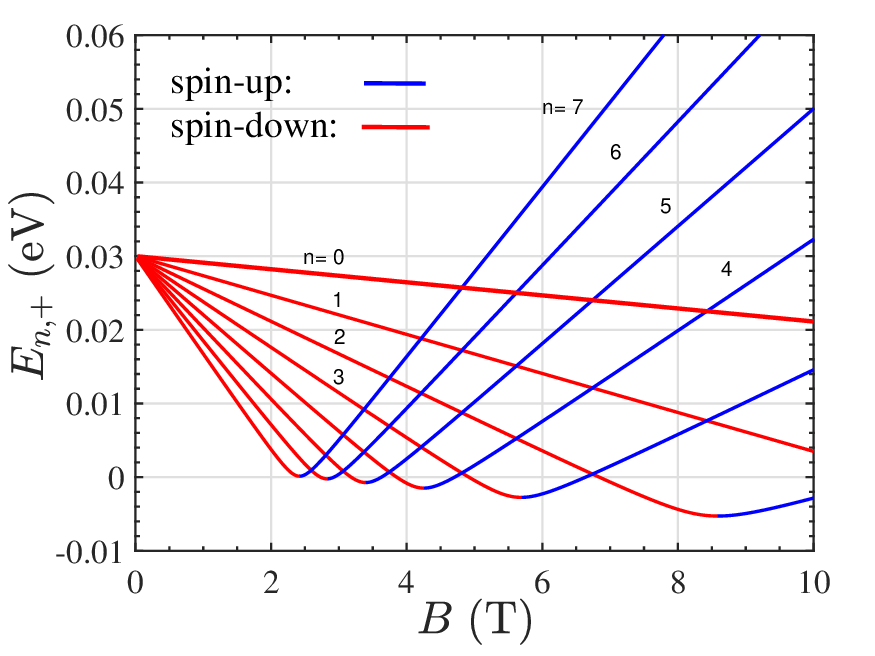}
\caption{The LL spectrum for first few $n$ are plotted with the magnetic field for the conduction band. The red and blue colors indicate the down and up out-of plane spin polarization, respectively. The same LL undergoes spin sign reversal upon a change in the magnetic field. Parameters taken are $\epsilon R^2= 0.03$ eV and $\alpha=10^{-11}$ eV-m}
\label{fig:landaulevels}
\end{figure}
\begin{figure}[t]
\centering
\includegraphics[width=0.90\linewidth]{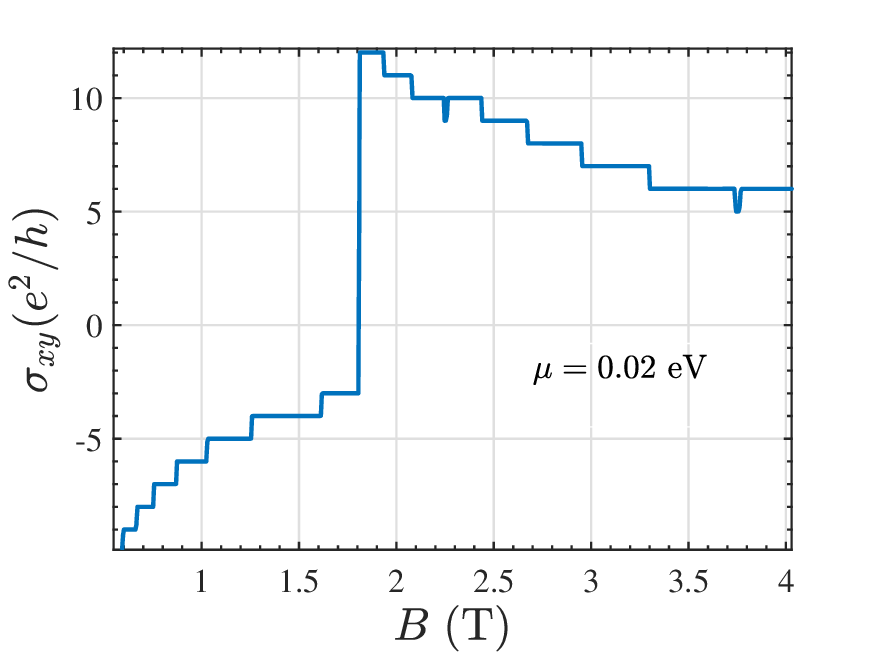}
\caption{The Hall conductivity is plotted the magnetic field $B(T)$ by keeping the chemical potential $\mu=0.02$ eV with other parameters same as in Fig.~(\ref{fig:landaulevels}).}
 \label{fig:hc}
\end{figure}
The Landau levels for $n>0$ are given as
\begin{equation}\label{eqn:lls}
E_{\xi}=-\frac{\hbar\omega_c}{2}+\lambda \sqrt{(n\hbar\omega_c-\epsilon R^2)^2+n\beta^2}
\end{equation}
where $\xi\equiv \{n,k_y,\lambda\}$ with $n=1,2,3\cdots$ being the LL indices. The corresponding eigenstates given as
\begin{equation}
\Psi_{\xi}(x,y)=\frac{e^{ik_yy}}{\sqrt{L_y}}
\begin{bmatrix} \phi_{n-1}(\tilde{x}) \\ c_n^\lambda ~\phi_{n}(\tilde{x})
\end{bmatrix},
\end{equation}
with $c_{n}^\lambda=\big[E_\xi+\hbar\omega_c/2-\sqrt{(E_\xi+\hbar\omega_c/2)^2-n\beta^2}\big]/(\sqrt{n}\beta)$ and $\phi_n(\tilde{x})=[1/\sqrt{2^{n} n! ~ l_c \sqrt{\pi}}~] ~\exp(-\tilde{x}^{2}/2)H_n(\tilde{x})$ is the usual one-dimensional quantum harmonic oscillator wave function in terms of the Hermite polynomial $H_n(\tilde{x})$. The lowest LL corresponding to $n=0$ is $E_0 = -\hbar\omega_c/2 $ independent of the RSOC with the corresponding ground state wave function as
\begin{equation}
\Psi_0(x,y)=\frac{e^{ik_yy}}{\sqrt{L_y}} \begin{bmatrix}
0\\ \phi_0(\tilde{x})   \end{bmatrix}.
\end{equation}
Similar to the previous case, here we find that the out-of-plane spin polarization in each  LL is $\langle \sigma_z\rangle\propto \lambda {\rm sgn}(n\hbar\omega_c-\epsilon R^2)$, indicating that the spin texture changes sign at the magnetic field when $n\hbar\omega_c=\epsilon R^2$. It is worth noting that for different LLs, the out-of-plane spin polarization changes sign at different values of the magnetic field. We note that only for the lowest LL ($n=0$) the spin texture remains unchanged as the magnetic field varies.

We now plot the spin-resolved LLs with magnetic field using  Eq.~(\ref{eqn:lls})  for the conduction band in Fig.~(\ref{fig:landaulevels}). The features of LLs are quite non-trivial, they exhibit a negative slope with the magnetic field, i.e., $dE_n/dB<0$ initially as long as $n\hbar\omega_c<\epsilon R^2$. However, after exhibiting a minima at the boundary $n\hbar\omega_c=\epsilon R^2$ the slopes become positive i.e., $dE_n/dB>0$. The out-of-plane spin texture also gets flipped across this boundary, as discussed earlier. The LLs with negative slopes are denoted by red lines, indicating spin-down branches, whereas the LLs with positive slopes in blue lines couple to spin-up branch. Another noticeable point is that for a negative slope $E_{n+1}<E_{n}$ whereas this order is reversed for a positive slope. Here we mention that such sign reversal of the slope generally occurs between two opposite bands, for example, LLs in graphene \cite{GusyninPRL} or dice-lattice \cite{xzfz-d19q}. Contrary to that, in our case such a sign reversal of the LL slope occurs in the same band by tuning the magnetic field.
\begin{figure*}[t]
\centering
\subfigure[]
{ \includegraphics[width=0.4\linewidth]{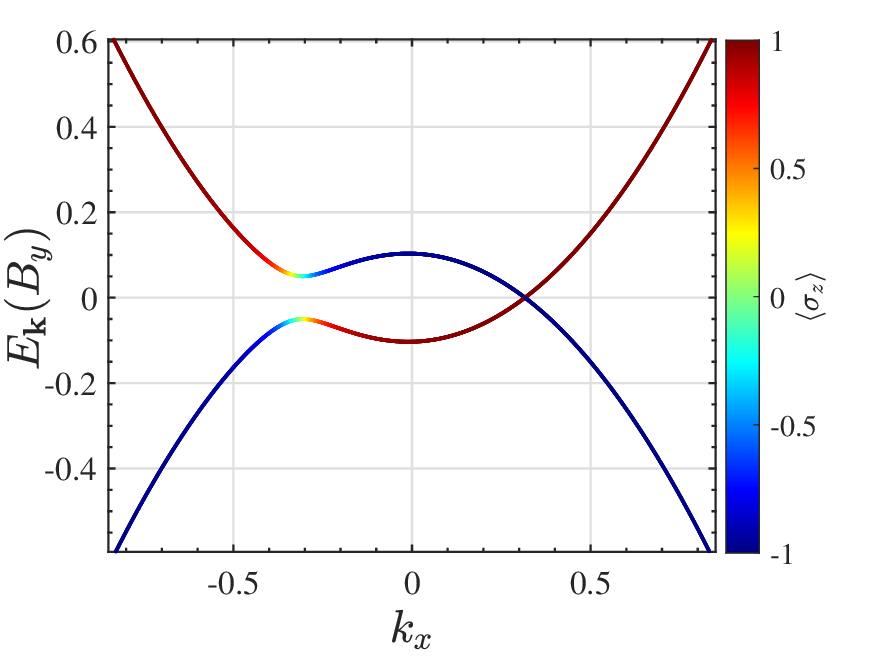} \label{fig:bs2}}
\centering
\subfigure[]
{\includegraphics[width=0.4\linewidth]{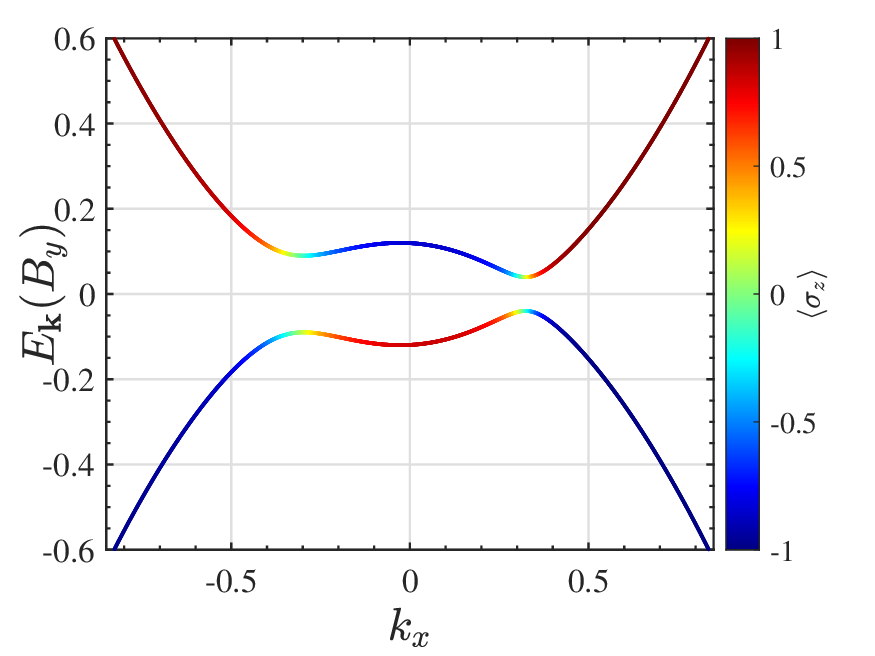} \label{fig:bs3}}
\centering

\label{fig:bs}
\caption{Energy spectrum of the system driven by the parallel magnetic field. Fig.~\ref{fig:bs2} shows the closing of the nodal gap at $(B_x,By)=(0,B_y^c)$. In Fig.~\ref{fig:bs3}, we show that the nodal gap opens beyond $B_y^c$ but with Chern number $C_{B}=0$, i.e., a trivial insulator. Energy and momentum are taken in the unit $\epsilon k_F^2$ and $k_F$ respectively. Parameters used are $R=0.3 $ and $\alpha =0.08$ in the units of $k_F$ and $\epsilon k_F$ respectively.}
\end{figure*}

Now with the exact LLs and Landau eigen states obtained above, we proceed to evaluate the quantum Hall conductivity using the Kubo-Greenwood formula based on the linear response theory} \cite{Charbonneau1982,Vasilopoulos2003,Tahir2016,Tutul2016,Aryan2026}     
\begin{equation} \label{eqn:kubo1}
\sigma_{xy}^{H} = \frac{ie^2\hbar}{A} \sum_{\xi \ne \xi'}(f_{\xi}- f_{\xi'}) \frac{\langle \Psi_{\xi} | \hat{v}_x | \Psi_\xi' \rangle \langle \Psi_\xi'| \hat{v}_y | \Psi_\xi\rangle}{(E_{\xi}- E_{\xi'})^2}.
\end{equation}
Here, $f_\xi=[\exp\{(E_\xi-\mu)/k_BT\}+1]^{-1}$ is the Fermi-Dirac distribution function at temperature $T$. The velocity operators can be obtained by using $\hat{v}_a=\hbar^{-1}\partial H/\partial k_a$ with $\{a=x,y\}$, as
\begin {equation}
\begin{aligned}
 \hat{v}_{x}= i\sqrt{2} \frac{\epsilon}{\hbar l_c}(\hat{a}^\dagger-\hat{a}) \sigma_z-\frac{\alpha}{\hbar} \sigma_y, \\
\hat{v}_{y}=\sqrt{2} \frac{\epsilon}{\hbar l_c}(\hat{a}^\dagger+\hat{a})\sigma_z +\frac{\alpha}{\hbar}\sigma_x,
\end{aligned}
\end{equation}

 This summation can be further simplied using  $\sum_{\xi,\xi'}\rightarrow~ [A/(2\pi l_c^2)]\sum_{\lambda\lambda'}\sum_{n,n'}$. Here, we used $\sum_{k_y}=\frac{L_y}{2\pi}\int_{0}^{L_x/l_c^2}dk_y$. Using these the Eq.~(\ref{eqn:kubo1}) reduces to
\begin{equation} \label{qhall}
\sigma_{xy}^{H}=\frac{i e^2\hbar}{2\pi l_c^2}\sum_{n,n'}\sum_{\lambda,\lambda'} \frac{ (f_{n,\lambda}-f_{n',\lambda'}) }{(E_{n,\lambda}-E_{n',\lambda'})^2 }\mathcal{M}_{n,n'}^{\lambda,\lambda'},
\end{equation}
where the velocity matrix product $ \mathcal{M}_{n,n'}^{\lambda,\lambda'}= \langle \Psi_{n}^\lambda| \hat{v}_x | \Psi_{n'}^{\lambda'} \rangle \langle \Psi_{n'}^{\lambda'}| \hat{v}_y | \Psi_{n}^\lambda\rangle$ is obtained as
\begin{align}
 \mathcal{M}_{nn'}^{\lambda\lambda'}= i\{ \mathcal{R}_{nn'}^{\lambda\lambda'}\delta_{n,n'+1} + \mathcal{S}_{nn'}^{\lambda\lambda'}\delta_{n,n'-1}\} \nonumber\\\times\{\mathcal{T}_{nn'}^{\lambda\lambda'}\delta_{n',n+1} + \mathcal{U}_{nn'}^{\lambda\lambda'}\delta_{n',n-1}\},
\end{align}
with corresponding elements $\mathcal{R}_{nn'}^{\lambda\lambda'},~\mathcal{S}_{nn'}^{\lambda\lambda'},~\mathcal{T}_{n'n}^{\lambda'\lambda} and ~\mathcal{U}_{n'n}^{\lambda'\lambda}$ given in the Appendix \ref{appen.}. The zeroth LL contribution to conductivity is obtained separately as 
\begin {equation} \label{qhallzero}
\sigma_{xy,0}^H= \frac{i e^2\hbar}{2\pi l_c}\bigg[\frac{ (f_0^\lambda-f_{1}^{\lambda'})~ \mathcal{M}_{01}^{\lambda,\lambda'}}{(E_{0,\lambda}-E_{1,\lambda'})^2}  + \frac{ (f_1^\lambda-f_{0}^{\lambda'})~ \mathcal{M}_{10}^{\lambda,\lambda'}}{(E_{1,\lambda}-E_{0,\lambda'})^2} \bigg].
\end{equation}

Here, the velocity matrix elements ${M}_{01}^{\lambda,\lambda'}$ and ${M}_{10}^{\lambda,\lambda'}$ are also given in  Appendix \ref{appen.}.
\begin{figure*}[t]
    \centering
     \subfigure[]
        {\includegraphics[width=0.45\linewidth]{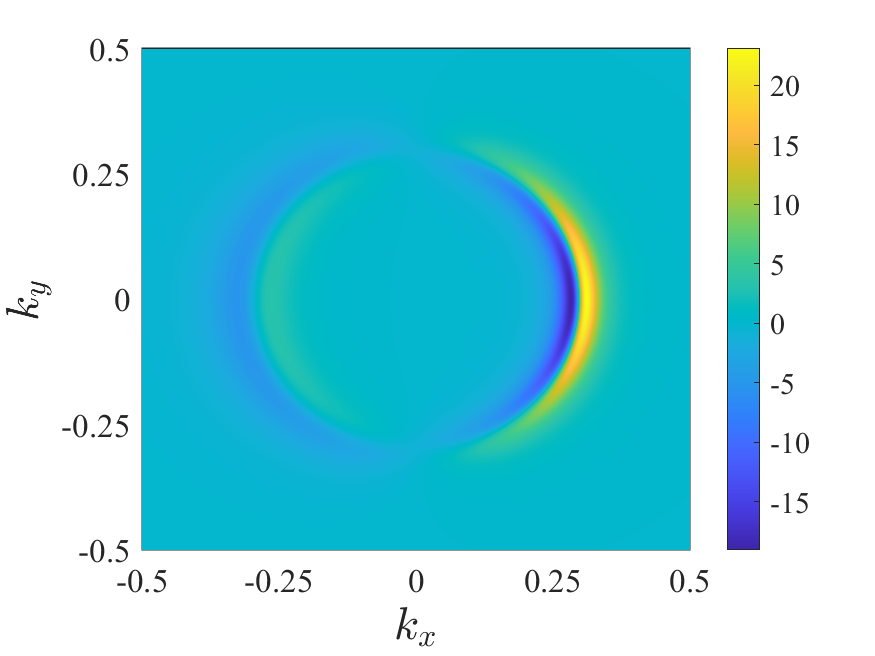} \label{fig:ddti}}
         \centering
      \subfigure[]
       { \includegraphics[width=0.45\linewidth]{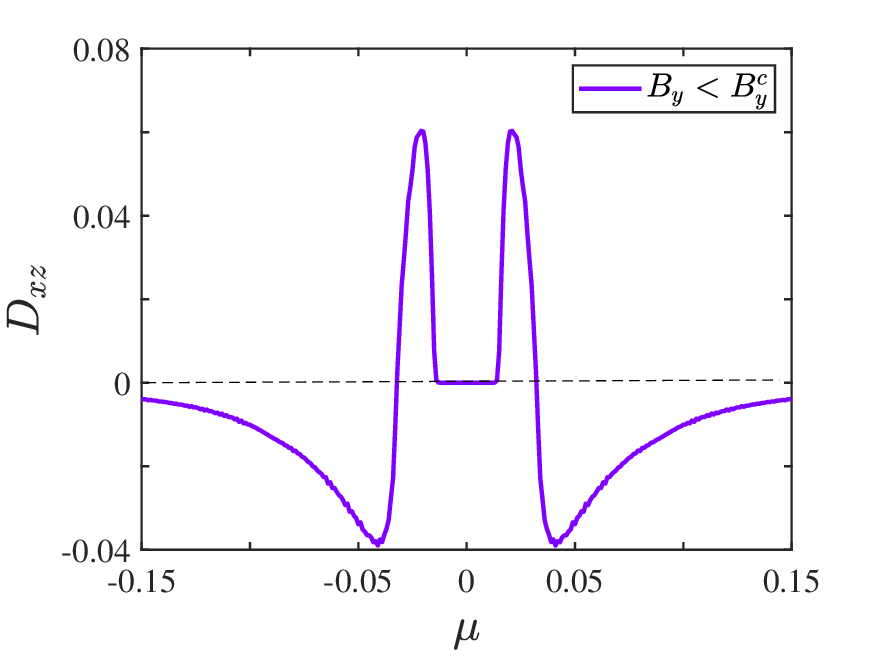} \label{fig:nlti}}
       \centering
       \subfigure[]
      {\includegraphics[width=0.45\linewidth]{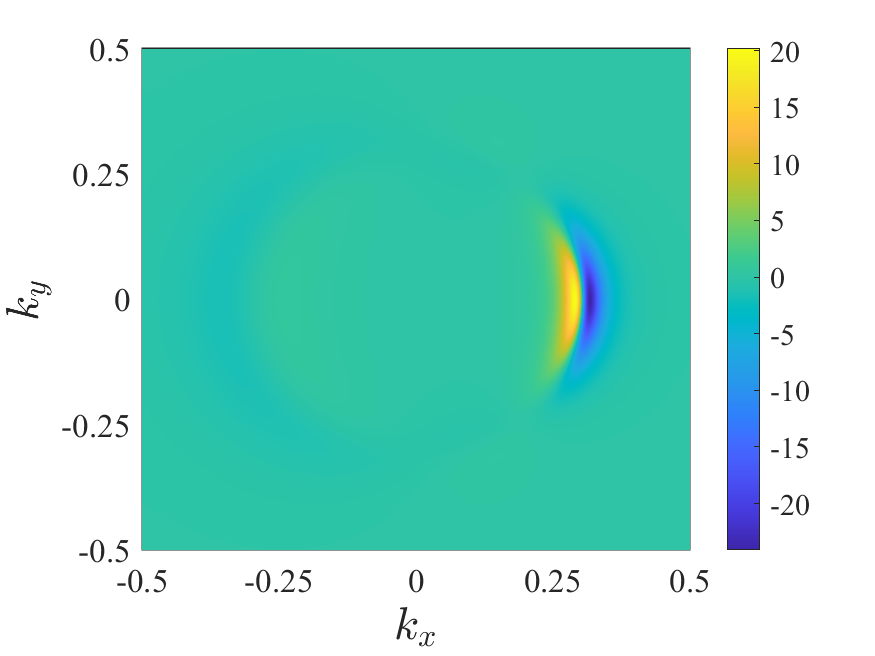} \label{fig:ddni}}
      \centering
       \subfigure[]
      {\includegraphics[width=0.45\linewidth]{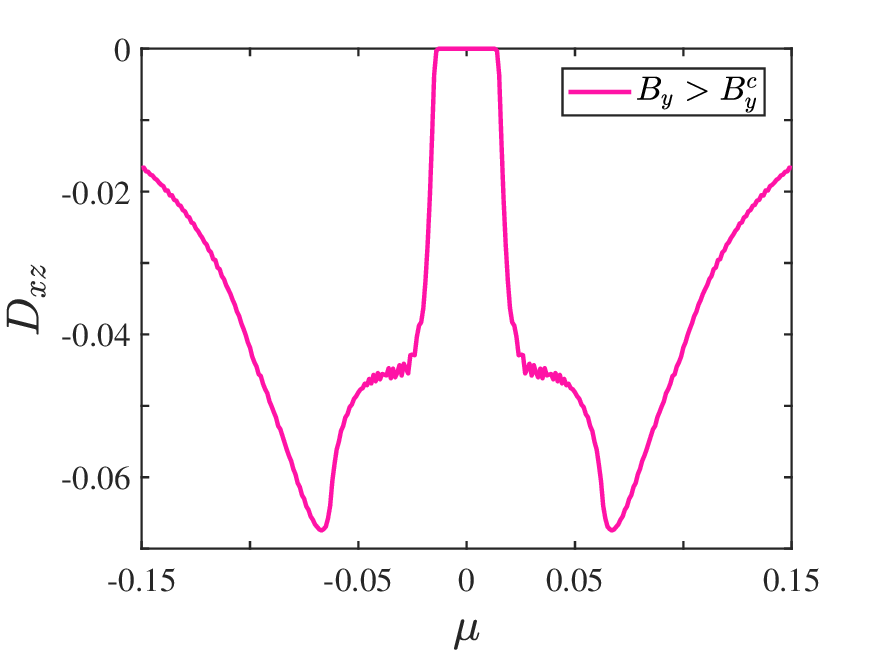} \label{fig:nlni}}

 \caption{ The Berry curvature density of the parallel field driven system for (a) Chern insulating phase and (c) trivial insulating phase are plotted in momentum space. The corresponding BCDs are plotted in  (b) and (d), respectively. We plot only $D_{xz}$ component of the Berry curvature dipole (BCD) as the other components are zero. Here, we normalize $v_x\Omega_z$ and $D_{xz}$ by $v_F k_F^{-2}$ and $k_F^{-1}$, respectively.}
    \label{fig:BCD} 
\end{figure*}
We evaluate the Hall conductivity using Eqs.~\eqref{qhall} and \eqref{qhallzero} and plot it as a function of the magnetic field in Fig.~(\ref{fig:hc}), which reveals a peculiar behavior directly reflecting the nature of the LLs. We note that quantum-Hall conductivity exhibits usual pattern of quantization with steps in units of $e^2/h$ corresponding to each Hall edge mode. At a low magnetic field, initially Hall conductivity is purely due to the LLs with a negative slope and a large number of LLs are occupied by the chemical potential. Hence, the Hall conductivity starts with a larger value. The negative sign of the Hall conductivity even for conduction band is non-trivial which is the direct manifestation of the negative slope of LLs. The negative sign can be further understood via Streda's formula of Hall conductivity \cite{Streda1982,RevModPhys821959}
\begin{eqnarray}
\sigma_{xy}^{H} &=& -e\left(\frac{\partial n_e}{\partial B}\right)_{\mu}
\nonumber\\
&=& -\frac{e^2}{h}\sum_{n,\lambda}\Theta(\mu-E_{n,\lambda})
+\frac{e^2B}{h}\sum_{n,\lambda}\delta(\mu-E_{n,\lambda})
\frac{\partial E_{n,\lambda}}{\partial B}
\nonumber
\end{eqnarray}
where $n_e=\sum_{\xi}\Theta(\mu-E_{\xi})$ is the carrier density with $\Theta(...)$ as the Heaviside step function. The above formula suggests that the sign of Hall conductivity is sensitive to the slope of the LLs. With the further increase of LLs there is a sudden discontinuity in the Hall conductivity with a big jump with changing sign. This can be attributed to the fact that the $\mu$ now intersects LLs with positive slope and higher $n$. Few deviation from the regular steps in the Hall conductivity in the second region are due to the contribution of few spin-down LLs (positive slope) with lower LL index $n$. It is quite unusual that the quantum-Hall conductivity gets a sign reversal without varying the chemical potential between two bands. As mentioned earlier, such sign reversal in Hall conductivity generally found in two-band Dirac semimetals--like graphene or the surface of $3$D topological insulator by varying the chemical potential between two bands. However, here such sign reversal occurs in the same band by tuning the magnetic field.

 \subsection{Parallel magnetic field driven nonlinear Hall effect}
 
Now we study the nonlinear Hall effect. As discussed previously, the nonlinear Hall response arises even in a system without breaking TRS but with broken IS. Our present system is already a magnetic nodal ring with a topological gap arising from intrinsically broken TRS as well as IS. However, because of the isotropic Berry curvature, the system does not exhibit the nonlinear Hall effect. In order induce an anisotropy in the Berry curvature we apply a parallel magnetic field ${\bf B}=(B_x,B_y,0)$ that directly couples with the spin of the system and resulting the new Hamiltonian as $  \mathcal{H}=  {\bf d}\cdot{\bf\sigma}+ g\mu_B(B_x  \sigma_x + B_y \sigma_y)$ with $g$ as Lande g-factor and $\mu_B$ as Bohr magnetron. We set $g\mu_B=1$ and obtain the corresponding energy as ${E}_{\mathbf{k},\lambda}(B_x,B_y)=\lambda\sqrt{\epsilon^2( k^2- R^2)^2 + (\alpha k_y+ B_x)^2 +(\alpha k_x - B_y)^2}$ which is anisotropic. We observe that this parallel field modifies the topological gap to $\Delta'=(\alpha R-B_y)$ at the node $(k_x,k_y)=(R,0)$. It is now clear that this gap can be easily closed at $B_y^c=\alpha R$, and with the further increase of the field, the gap can be reopened. The closing and reopening of the gap are shown in the modified band anisotropic band structure in Fig.~ \ref{fig:bs2}. Such band anisotropy and gap closing, driven by a parallel magnetic field, have recently been considered in the context of the Josephson effect by W.-T. Liu et al. \cite{josephson2025}. In order to further investigate the topological nature of the gap on both sides of the closing point, we calculate its  Berry curvature and Chern number as follows.
The corresponding Berry curvature can be immediately obtained as
\begin{align}
\Omega_z^{k}(B) = \frac{\alpha^2 \epsilon( k^2+R^2) + 2 \epsilon \alpha(B_xk_y- B_y k_x)}{2\left| \mathcal{E}_{k,\lambda} \right|^3}.
\end{align}
The Chern number switches from $C_{B}=1$ to $0$ for $B_y<B_y^c$ and $B_y>B_y^c$, respectively. Hence, we find a reverse topological phase transition by applying a parallel magnetic field. However, note that in both phases the Berry curvature is anisotropic that stems from the parallel field-driven anisotropic energy spectrum. Next, we aim to evaluate the nonlinear Hall conductivity in both phases.

When $B_y<B_y^c$, the system is already a Chern insulator exhibiting linear anomalous Hall effect. We calculate the second order Hall response arising from the BCD using Eq.~(\ref{eqn:bcd}) where the different velocity components are used as  ${v}_{x_B}^\lambda=[2\epsilon^2 k_x( k^2-R^2)+\alpha(\alpha k_x-B_y)]/\mathcal{E}_{k,\lambda}$ and $v_{y_B}^\lambda=v_{y_1}$.  We observe that the Berry curvature dipole density (${v}_{x_B}\Omega_B$) is no longer odd in momentum and contributes to the finite BCD $D_{xz}$. The other component $D_{yz}$ still remains zero as the product of the corresponding velocity component and the Berry curvature is odd in momentum. Additionally, when $B_y>B_y^c$, i.e., the system is just a trivial insulator, the nonlinear Hall still survives because of the anisotropic Berry curvature. We plot the Berry curvature dipole density  in momentum space, and corresponding BCD component with the chemical potential, in the Fig.~(\ref{fig:BCD}) for both the phases. The dipole density is clearly showing an asymmetric nature in momentum space enabling the non zero nonlinear Hall response quantified by the BCD. The BCD as a function of chemical potential for topological phase is shown in the Fig.~\ref{fig:nlti}, where BCD exhibits both sign with sharp peaks. In a similar manner, corresponding to the trivial insulator phase ($B_y>B_y^c$) phase, the anisotropic dipole density and its corresponding BCD are shown in Fig.~\ref{fig:ddni} and Fig.~\ref{fig:nlni}, respectively. It can be seen in Fig.~\ref{fig:nlni} that no sign flip occurs in the peak position for such a phase.

We comment here that contrary to the linear Hall effect, where the Hall conductivity is an inter-band phenomenon, the non linear Hall is purely intra-band. As a result, it remains persistent irrespective of the nature of the gap, i.e., whether the gap is topological or insulating, is not relevant. It can be seen from the plot of BCD in Fig.~\ref{fig:BCD}(a), which shows that the nonlinear begins to increase as the chemical potential penetrates inside the bulk, whereas the anomalous Hall is restricted as long as the chemical potential resides inside the topological gap. The topological nature and corresponding Hall response of the system for the various parallel-field regimes are briefly summarised in the TABLE~\ref{tab:hall_response}.
\begin{table}[htpb]
\caption{Linear and nonlinear Hall responses in different in-plane magnetic field regimes.}
\label{tab:hall_response}
\begin{ruledtabular}
\renewcommand{\arraystretch}{1.15}
\begin{tabular}{|l|c|c|c|}
Order of Hall effect & $B_y=0$ & $B_y<B_y^c$ & $B_y>B_y^c$ \\
\hline
Intrinsic linear Hall    & $\checkmark$ & $\checkmark$ & $\times$ \\
Nonlinear Hall & $\times$     & $\checkmark$ & $\checkmark$ \\
\end{tabular}
\end{ruledtabular}
\end{table}

\section{ Summary}\label{summary}
We have studied the linear and nonlinear Hall effect in the RSOC two-dimensional nodal ring. We first show that although the RSOC is the result of breaking IS, but along with the TRS breaking term it can open up a topological gap leading to a linear anomalous Hall phase. Here, we also studied the perpendicular magnetic field-induced integer quantum Hall effect as another route to the linear Hall effect. We obtained exact Landau levels and subsequently used the Kubo-Greenwood formula to obtain the quantum Hall quantization. We showed that the Hall conductivity exhibits a sign reversal with varying magnetic field in the same band. This is because of the sign reversal of LL slope with the magnetic field across the ring boundary. Most interestingly, we also found that applying a parallel magnetic field can close and reopen the topological gap, thereby causing a topological phase transition from a nontrivial topological phase to a trivial insulating phase. However, the parallel field also induced a band anisotropy resulting in the emergence of the nonlinear Hall effect. The nonlinear Hall effect remains robust across both phases. It is also observed that BCD as a function of chemical potential shows a sign-changing peak in the topological phase, whereas in the insulating phase such peak remains single-signed. Therefore, such contrasting behavior of nonlinear Hall in two different phases can serve as a probe for studying such topological phase transitions.

\section{Acknowledgement}
SK Firoz Islam acknowledges financial support from the project ANRF/ECRG/2024/005166/PMS.
\section{Data Availability}
The data that supports the findings of this article are not publicly available. The data are available from the authors upon reasonable request.

\appendix 
\section{ Matrix elements of the velocity operator $\hat{v}_x$ and $\hat{v}_y$ for $n>0$ and $n=0$ }\label{appen.} 
The coefficents $\mathcal{R}_{nn'}^{\lambda\lambda'},~\mathcal{S}_{nn'}^{\lambda\lambda'},~ \mathcal{T}_{nn'}^{\lambda,\lambda'}$ and $\mathcal{U}_{nn'}^{\lambda,\lambda'}$ of the matrix product $\mathcal{M}_{n,n'}^{\lambda,\lambda'}$ for $n>0$ are given as
\begin{align}
\mathcal{R}_{nn'}^{\lambda\lambda'} 
=  \frac{1}{\hbar} \Big( \sqrt{2} \frac{\epsilon}{l_c}   \sqrt{n'} + \alpha  c_{n'}^{\lambda'} -  \sqrt{2} \frac{\epsilon}{l_c}  (c_{n}^{\lambda})^* c_{n'}^{\lambda'} \sqrt{n'+1} \Big)
 \end{align}
\begin{align}
\mathcal{S}_{nn'}^{\lambda\lambda'} = \frac{1}{\hbar}\Big( - \sqrt{2} \frac{\epsilon}{l_c}   \sqrt{n'-1} - \alpha c_{n}^{\lambda}  +  \sqrt{2} \frac{\epsilon}{l_c} c_{n}^{\lambda} c_{n'}^{\lambda'} \sqrt{n'} \Big) 
\end{align}

\begin{align}
\mathcal{T}_{n'n}^{\lambda'\lambda}=   \frac{1}{\hbar}\big(\sqrt{2} \frac{\epsilon}{l_c}  \sqrt{n} + \alpha  c_{n}^{\lambda} - \sqrt{2} \frac{\epsilon}{l_c}  \, (c_{n'}^{\lambda'})^* c_{n}^{\lambda}) \sqrt{n+1} \,big)
 \end{align}
\begin{align}
 \mathcal{U}_{n'n}^{\lambda'\lambda}=  \frac{1}{\hbar}\big(\sqrt{2} \frac{\epsilon}{l_c}   \sqrt{n-1} + \alpha  (c_{n'}^{\lambda'})^* -  \sqrt{2} \frac{\epsilon}{l_c}  (c_{n'}^{\lambda'})^* c_{n}^{\lambda}) \sqrt{n}\big) 
\end{align}
Also, the matrix elements $\mathcal{M}_{01}^{\lambda\lambda'}= \langle \Psi_{0}^\lambda| \hat{v}_x | \Psi_{1}^{\lambda'} \rangle \langle \Psi_{1}^{\lambda'}| \hat{v}_y | \Psi_{0}^\lambda\rangle$ and $\mathcal{M}_{10}^{\lambda\lambda'}= \langle \Psi_{1}^\lambda| \hat{v}_x | \Psi_{0}^{\lambda'} \rangle \langle \Psi_{0}^{\lambda'}| \hat{v}_y | \Psi_{1}^\lambda\rangle$
are given as 
\begin{align}
\mathcal{M}_{01}^{\lambda\lambda'}=\frac{-i}{\hbar} 
\bigg [\alpha^2 + \bigg(\sqrt{2} \frac{\epsilon}{l_c}\bigg)^2 (c_{1}^{\lambda'})^*c_1^{\lambda}  -  \sqrt{2} \frac{\epsilon}{l_c} \alpha[ (c_1^{\lambda})^*+(c_1]\bigg]
\end{align}
\begin{align}
\mathcal{M}_{10}^{\lambda\lambda'}=\frac{i}{\hbar}
\bigg [\alpha^2 + \bigg(\sqrt{2} \frac{\epsilon}{l_c}\bigg)^2 (c_{1}^{\lambda'})^*c_1^{\lambda} -  \sqrt{2} \frac{\epsilon}{l_c} \alpha [ (c_1^{\lambda})^*+(c_1]\bigg]
\end{align}
\bibliography{nhall}

@article{74kyd71n,
  title = {Quantized Spin Hall Effect in Three-Dimensional Nodal-Ring Semimetal: Geometric Scaling and Symmetry-Engineered Spin Response},
  author = {Chen, Jiali and Cui, Chaoxi and Yu, Zhi-Ming and Jiang, Wei and Yao, Yugui},
  journal = {Phys. Rev. Lett.},
  volume = {137},
  issue = {8},
  pages = {086301},
  numpages = {8},
  year = {2026},
  month = {Aug},
  publisher = {American Physical Society},
  doi = {10.1103/74ky-d71n},
  url = {https://link.aps.org/doi/10.1103/74ky-d71n}
}

@article{josephson2025 ,
  title = {Controllable Josephson diode effect, $0\text{\ensuremath{-}}\ensuremath{\pi}$ transition, and switch effect in superconductor/two-dimensional Weyl nodal line semimetal/superconductor junctions},
  author = {Liu, Wen-Ting and Zhao, Shu-Chang and Cheng, Qiang and Sun, Qing-Feng},
  journal = {Phys. Rev. B},
  volume = {112},
  issue = {5},
  pages = {054509},
  numpages = {10},
  year = {2025},
  month = {Aug},
  publisher = {American Physical Society},
  doi = {10.1103/tlqg-268j},
  url = {https://link.aps.org/doi/10.1103/tlqg-268j}
}

@article{Torma2023,
  title = {Essay: Where Can Quantum Geometry Lead Us?},
  author = {T\"orm\"a, P\"aivi},
  journal = {Phys. Rev. Lett.},
  volume = {131},
  issue = {24},
  pages = {240001},
  numpages = {7},
  year = {2023},
  month = {Dec},
  publisher = {American Physical Society},
  doi = {10.1103/PhysRevLett.131.240001},
  url = {https://link.aps.org/doi/10.1103/PhysRevLett.131.240001}
}

@article{Yu2025,
 title     = {Quantum geometry in quantum materials},
  author    = {Yu, Jiabin and Bernevig, B. Andrei and Queiroz, Raquel and Rossi, Enrico and T{\"o}rm{\"a}, P{\"a}ivi and Yang, Bohm-Jung},
  journal   = {npj Quantum Materials},
  year      = {2025},
  volume    = {10},
  number    = {1},
  pages     = {101},
  month     = {oct},
  issn      = {2397-4648},
  doi       = {10.1038/s41535-025-00801-3},
  url       = {https://doi.org/10.1038/s41535-025-00801-3},
}

@article{RevModPhys821959,
  title = {Berry phase effects on electronic properties},
  author = {Xiao, Di and Chang, Ming-Che and Niu, Qian},
  journal = {Rev. Mod. Phys.},
  volume = {82},
  issue = {3},
  pages = {1959--2007},
  numpages = {0},
  year = {2010},
  month = {Jul},
  publisher = {American Physical Society},
  doi = {10.1103/RevModPhys.82.1959},
  url = {https://link.aps.org/doi/10.1103/RevModPhys.82.1959}
}

@article{Haldane2004,
  title = {Berry Curvature on the Fermi Surface: Anomalous Hall Effect as a Topological Fermi-Liquid Property},
  author = {Haldane, F. D. M.},
  journal = {Phys. Rev. Lett.},
  volume = {93},
  issue = {20},
  pages = {206602},
  numpages = {4},
  year = {2004},
  month = {Nov},
  publisher = {American Physical Society},
  doi = {10.1103/PhysRevLett.93.206602},
  url = {https://link.aps.org/doi/10.1103/PhysRevLett.93.206602}
}

@article{sodemann2015,
  title = {Quantum Nonlinear Hall Effect Induced by Berry Curvature Dipole in Time-Reversal Invariant Materials},
  author = {Sodemann, Inti and Fu, Liang},
  journal = {Phys. Rev. Lett.},
  volume = {115},
  issue = {21},
  pages = {216806},
  numpages = {5},
  year = {2015},
  month = {Nov},
  publisher = {American Physical Society},
  doi = {10.1103/PhysRevLett.115.216806},
  url = {https://link.aps.org/doi/10.1103/PhysRevLett.115.216806}
}

@article{Ortix2021,
  author  = {Carmine Ortix},
  title   = {Nonlinear Hall Effect with Time-Reversal Symmetry: Theory and Material Realizations},
  journal = {Adv. Quantum Technol.},
  volume  = {4},
  number  = {9},
  pages   = {2100056},
  year    = {2021},
  doi     = {10.1002/qute.202100056}
}

@article{Du2021,
  author  = {Z. Z. Du and Hai-Zhou Lu and X. C. Xie},
  title   = {Nonlinear Hall effects},
  journal = {Nat. Rev. Phys.},
  volume  = {3},
  number  = {11},
  pages   = {744--752},
  year    = {2021},
  doi     = {10.1038/s42254-021-00359-6}
}

@article{Bandyopadhyay2024,
  author  = {Arka Bandyopadhyay and Nesta Benno Joseph and Awadhesh Narayan},
  title   = {Non-linear Hall effects: Mechanisms and materials},
  journal = {Mater. Today Electron.},
  volume  = {8},
  pages   = {100101},
  year    = {2024},
  doi     = {10.1016/j.mtelec.2024.100101}
}

@article{Klitzing1980,
  author  = {K. von Klitzing and G. Dorda and M. Pepper},
  title   = {New Method for High-Accuracy Determination of the Fine-Structure Constant Based on Quantized Hall Resistance},
  journal = {Phys. Rev. Lett.},
  volume  = {45},
  pages   = {494--497},
  year    = {1980},
  doi     = {10.1103/PhysRevLett.45.494}
}

@article{You2018,
  title = {Berry curvature dipole current in the transition metal dichalcogenides family},
  author = {You, Jhih-Shih and Fang, Shiang and Xu, Su-Yang and Kaxiras, Efthimios and Low, Tony},
  journal = {Phys. Rev. B},
  volume = {98},
  issue = {12},
  pages = {121109(R)},
  numpages = {6},
  year = {2018},
  month = {Sep},
  publisher = {American Physical Society},
  doi = {10.1103/PhysRevB.98.121109},
  url = {https://link.aps.org/doi/10.1103/PhysRevB.98.121109}
}

@article{Joseph2021,
doi = {10.1088/2053-1591/ac440b},
url = {https://doi.org/10.1088/2053-1591/ac440b},
year = {2021},
month = {dec},
publisher = {IOP Publishing},
volume = {8},
number = {12},
pages = {124001},
author = {Joseph, Nesta Benno and Roy, Saswata and Narayan, Awadhesh},
title = {Tunable topology and berry curvature dipole in transition metal dichalcogenide Janus monolayers},
journal = {Mater. Res. Express},
}

@article{Zhang2018,
  title = {Berry curvature dipole in Weyl semimetal materials: An ab initio study},
  author = {Zhang, Yang and Sun, Yan and Yan, Binghai},
  journal = {Phys. Rev. B},
  volume = {97},
  issue = {4},
  pages = {041101(R)},
  numpages = {6},
  year = {2018},
  month = {Jan},
  publisher = {American Physical Society},
  doi = {10.1103/PhysRevB.97.041101},
  url = {https://link.aps.org/doi/10.1103/PhysRevB.97.041101}
}

@article{Chuanchang2021,
  title = {Nonlinear transport in Weyl semimetals induced by Berry curvature dipole},
  author = {Zeng, Chuanchang and Nandy, Snehasish and Tewari, Sumanta},
  journal = {Phys. Rev. B},
  volume = {103},
  issue = {24},
  pages = {245119},
  numpages = {12},
  year = {2021},
  month = {Jun},
  publisher = {American Physical Society},
  doi = {10.1103/PhysRevB.103.245119},
  url = {https://link.aps.org/doi/10.1103/PhysRevB.103.245119}
}

@article{Xu2018,
  author  = {Xu, Su-Yang and Ma, Qiong and Shen, Huitao and Fatemi, Valla and Wu, Sanfeng and Chang, Tay-Rong and Chang, Guoqing and Mier Valdivia, Andr{\'e}s M. M. and Chan, Ching-Kit and Gibson, et al.},
  title   = {Electrically switchable Berry curvature dipole in the monolayer topological insulator WTe2},
  journal = {Nat. Phys.},
  volume  = {14},
  pages   = {900--906},
  year    = {2018},
  doi     = {10.1038/s41567-018-0189-6}
}

@article{Ma2019,
  author = {Ma, Qiong and
            Xu, Su-Yang and
            Shen, Huitao and
            MacNeill, David and
            Fatemi, Valla and
            Chang, Tay-Rong and
            Mier Valdivia, Andr\'es M. and
            Wu, Sanfeng and
            Du, Zongzheng and
            Hsu, Chuang-Han and
            others},
  title = {Observation of the Nonlinear Hall Effect under Time-Reversal-Symmetric Conditions},
  journal = {Nat. Phys.},
  volume = {565},
  number = {7739},
  pages = {337--342},
  year = {2019},
  doi = {10.1038/s41586-018-0807-6}
}

@article{Kang2019,
  author   = {Kang, Kaifei and Li, Tingxin and Sohn, Egon and Shan, Jie and Mak, Kin Fai},
  title    = {Nonlinear anomalous Hall effect in few-layer {WTe2}},
  journal  = {Nat. Matt.},
  year     = {2019},
  month    = apr,
  volume   = {18},
  number   = {4},
  pages    = {324--328},
  issn     = {1476-4660},
  doi      = {10.1038/s41563-019-0294-7},
  url      = {https://doi.org/10.1038/s41563-019-0294-7}
}

@article{Ye2023,
  title = {Control over Berry Curvature Dipole with Electric Field in ${\mathrm{WTe}}_{2}$},
  author = {Ye, Xing-Guo and Liu, Huiying and Zhu, Peng-Fei and Xu, Wen-Zheng and Yang, Shengyuan A. and Shang, Nianze and Liu, Kaihui and Liao, Zhi-Min},
  journal = {Phys. Rev. Lett.},
  volume = {130},
  issue = {1},
  pages = {016301},
  numpages = {6},
  year = {2023},
  month = {Jan},
  publisher = {American Physical Society},
  doi = {10.1103/PhysRevLett.130.016301},
  url = {https://link.aps.org/doi/10.1103/PhysRevLett.130.016301}
}

@Article{Araki2018,
author={Araki, Yasufumi},
title={Strain-induced nonlinear spin Hall effect in topological Dirac semimetal},
journal={Sci. Rep.},
year={2018},
month={Oct},
day={15},
volume={8},
number={1},
pages={15236},
issn={2045-2322},
doi={10.1038/s41598-018-33655-w},
url={https://doi.org/10.1038/s41598-018-33655-w}
}

@article{Chakraborty2022-bw,
  title     = {Nonlinear anomalous Hall effects probe topological
               phase-transitions in twisted double bilayer graphene},
  author    = {Chakraborty, Atasi and Das, Kamal and Sinha, Subhajit and Adak,
               Pratap Chandra and Deshmukh, Mandar M and Agarwal, Amit},
  journal   = {2D Mater.},
  publisher = {IOP Publishing},
  volume    =  {9},
  number    =  {4},
  pages     = {045020},
  month     =  {oct},
  year      =  {2022},
  doi={10.1088/2053-1583/ac8b93},
  url={https://iopscience.iop.org/article/10.1088/2053-1583/ac8b93}
}

@article{Qin2024,
  title = {Light-enhanced nonlinear Hall effect},
  author = {Qin, Fang and Chen, Rui and Lee, Ching Hua},
  journal = {Commun. Phys.},
  volume = {7},
  number = {1},
  pages = {368},
  year = {2024},
  month = {Nov},
  doi = {10.1038/s42005-024-01820-5},
  url = {https://doi.org/10.1038/s42005-024-01820-5}
}

@article{Xiao2010,
  title = {Berry phase effects on electronic properties},
  author = {Xiao, Di and Chang, Ming-Che and Niu, Qian},
  journal = {Rev. Mod. Phys.},
  volume = {82},
  issue = {3},
  pages = {1959--2007},
  numpages = {0},
  year = {2010},
  month = {Jul},
  publisher = {American Physical Society},
  doi = {10.1103/RevModPhys.82.1959},
  url = {https://link.aps.org/doi/10.1103/RevModPhys.82.1959}
}

@article{Nagaosa,
  title = {Anomalous Hall effect},
  author = {Nagaosa, Naoto and Sinova, Jairo and Onoda, Shigeki and MacDonald, A. H. and Ong, N. P.},
  journal = {Rev. Mod. Phys.},
  volume = {82},
  issue = {2},
  pages = {1539--1592},
  numpages = {0},
  year = {2010},
  month = {May},
  publisher = {American Physical Society},
  doi = {10.1103/RevModPhys.82.1539},
  url = {https://link.aps.org/doi/10.1103/RevModPhys.82.1539}
}

@article{Chen2022,
  title = {Photon-modulated linear and nonlinear anomalous Hall effects in type-II semi-Dirac semimetals},
  title = {Photon-modulated linear and nonlinear anomalous Hall effects in type-II semi-Dirac semimetals},
  author = {Chen, Jin-Na and Yang, Yan-Yan and Zhou, Yong-Long and Wu, Yong-Jia and Duan, Hou-Jian and Deng, Ming-Xun and Wang, Rui-Qiang},
  journal = {Phys. Rev. B},
  volume = {105},
  issue = {8},
  pages = {085124},
  numpages = {8},
  year = {2022},
  month = {Feb},
  publisher = {American Physical Society},
  doi = {10.1103/PhysRevB.105.085124},
  url = {https://link.aps.org/doi/10.1103/PhysRevB.105.085124}
}

@article{zhu2026,
  title = {Nonlinear Hall effect induced by two-frequency drives},
  author = {Zhu, Jiong-Yi and Chen, Rui and Zhou, Bin},
  journal = {Phys. Rev. B},
  volume = {113},
  issue = {19},
  pages = {195422},
  numpages = {19},
  year = {2026},
  month = {May},
  publisher = {American Physical Society},
  doi = {10.1103/17rp-cx9f},
  url = {https://link.aps.org/doi/10.1103/17rp-cx9f}
}

@article{Saha2023,
doi = {10.1088/1361-648X/acf1eb},
url = {https://doi.org/10.1088/1361-648X/acf1eb},
year = {2023},
month = {sep},
publisher = {IOP Publishing},
volume = {35},
number = {48},
pages = {485301},
author = {Saha, Soumadeep and Narayan, Awadhesh},
title = {Nonlinear Hall effect in Rashba systems with hexagonal warping},
journal = {J. Phys.: Condens. Matter},
}

@article{ankita2025,
  title = {Electric field induced second-order anomalous Hall transport in unconventional Rashba systems},
  author = {Bhattacharya, Ankita and Black-Schaffer, Annica M.},
  journal = {Phys. Rev. B},
  volume = {111},
  issue = {4},
  pages = {L041202},
  numpages = {8},
  year = {2025},
  month = {Jan},
  publisher = {American Physical Society},
  doi = {10.1103/PhysRevB.111.L041202},
  url = {https://link.aps.org/doi/10.1103/PhysRevB.111.L041202}

}

@article{PhysRevLett.123.116401,
  title = {Discovery of Weyl Nodal Lines in a Single-Layer Ferromagnet},
  author = {Feng, Baojie and Zhang, Run-Wu and Feng, Ya and Fu, Botao and Wu, Shilong and Miyamoto, Koji and He, Shaolong and Chen, Lan and Wu, Kehui and Shimada, Kenya and Okuda, Taichi and Yao, Yugui},
  journal = {Phys. Rev. Lett.},
  volume = {123},
  issue = {11},
  pages = {116401},
  numpages = {6},
  year = {2019},
  month = {Sep},
  publisher = {American Physical Society},
  doi = {10.1103/PhysRevLett.123.116401},
  url = {https://link.aps.org/doi/10.1103/PhysRevLett.123.116401}
}

@article{PhysRevB.102.125118,
  title = {Two-dimensional Weyl nodal-line semimetal in a ${d}^{0}$ ferromagnetic ${\mathrm{K}}_{2}\mathrm{N}$ monolayer with a high Curie temperature},
  author = {Jin, Lei and Zhang, Xiaoming and Liu, Ying and Dai, Xuefang and Shen, Xunan and Wang, Liying and Liu, Guodong},
  journal = {Phys. Rev. B},
  volume = {102},
  issue = {12},
  pages = {125118},
  numpages = {8},
  year = {2020},
  month = {Sep},
  publisher = {American Physical Society},
  doi = {10.1103/PhysRevB.102.125118},
  url = {https://link.aps.org/doi/10.1103/PhysRevB.102.125118}
}

@article{PhysRevB.99.035125,
  title = {Topological nodal-line semimetals in ferromagnetic rare-earth-metal monohalides},
  author = {Nie, Simin and Weng, Hongming and Prinz, Fritz B.},
  journal = {Phys. Rev. B},
  volume = {99},
  issue = {3},
  pages = {035125},
  numpages = {9},
  year = {2019},
  month = {Jan},
  publisher = {American Physical Society},
  doi = {10.1103/PhysRevB.99.035125},
  url = {https://link.aps.org/doi/10.1103/PhysRevB.99.035125}
}

@article{PhysRevB.71.085315,
  title = {Effect of in-plane magnetic field on the spin Hall effect in a Rashba-Dresselhaus system},
  author = {Chang, Ming-Che},
  journal = {Phys. Rev. B},
  volume = {71},
  issue = {8},
  pages = {085315},
  numpages = {5},
  year = {2005},
  month = {Feb},
  publisher = {American Physical Society},
  doi = {10.1103/PhysRevB.71.085315},
  url = {https://link.aps.org/doi/10.1103/PhysRevB.71.085315}
}

@article{PhysRevB.83.245428,
  title = {Parallel magnetic field driven quantum phase transition in a thin topological insulator film},
  author = {Zyuzin, A. A. and Hook, M. D. and Burkov, A. A.},
  journal = {Phys. Rev. B},
  volume = {83},
  issue = {24},
  pages = {245428},
  numpages = {4},
  year = {2011},
  month = {Jun},
  publisher = {American Physical Society},
  doi = {10.1103/PhysRevB.83.245428},
  url = {https://link.aps.org/doi/10.1103/PhysRevB.83.245428}
}

@article{Taskin2017,
  author  = {Taskin, A. A. and Legg, Henry F. and Yang, Fan and Sasaki, Satoshi and Kanai, Yasushi and Matsumoto, Kazuhiko and Rosch, Achim and Ando, Yoichi},
  title   = {Planar Hall effect from the surface of topological insulators},
  journal = {Nature Communications},
  year    = {2017},
  volume  = {8},
  pages   = {1340},
  doi     = {10.1038/s41467-017-01474-8}
}

@book{Winkler2003,
  author    = {Winkler, R.},
  title     = {Spin-Orbit Coupling Effects in Two-Dimensional Electron and Hole Systems},
  series    = {Springer Tracts in Modern Physics},
  volume    = {191},
  year      = {2003},
  publisher = {Springer Berlin Heidelberg},
  address   = {Berlin, Heidelberg},
  doi       = {10.1007/b13586},
  url       = {https://doi.org/10.1007/b13586}
}

@article{RevModPhys.87.1213,
  title = {Spin Hall effects},
  author = {Sinova, Jairo and Valenzuela, Sergio O. and Wunderlich, J. and Back, C. H. and Jungwirth, T.},
  journal = {Rev. Mod. Phys.},
  volume = {87},
  issue = {4},
  pages = {1213--1260},
  numpages = {47},
  year = {2015},
  month = {Oct},
  publisher = {American Physical Society},
  doi = {10.1103/RevModPhys.87.1213},
  url = {https://link.aps.org/doi/10.1103/RevModPhys.87.1213}
}

@article{Charbonneau1982,
  author    = {Charbonneau, M. and van Vliet, K. M. and Vasilopoulos, P.},
  title     = {Linear response theory revisited {III}: {O}ne-body response formulas and generalized {B}oltzmann equations},
  journal   = {J. Math. Phys.},
  volume    = {23},
  number    = {2},
  pages     = {318--336},
  year      = {1982},
  month     = {feb},
  publisher = {American Institute of Physics},
  doi       = {10.1063/1.525355},
  url       = {https://doi.org/10.1063/1.525355}
}

@article{Vasilopoulos2003,
  title = {Magnetotransport in a two-dimensional electron gas in the presence of spin-orbit interaction},
  author = {Wang, X. F. and Vasilopoulos, P.},
  journal = {Phys. Rev. B},
  volume = {67},
  issue = {8},
  pages = {085313},
  numpages = {7},
  year = {2003},
  month = {Feb},
  publisher = {American Physical Society},
  doi = {10.1103/PhysRevB.67.085313},
  url = {https://link.aps.org/doi/10.1103/PhysRevB.67.085313}
}

@article{Tahir2016,
  title = {Quantum magnetotransport properties of a ${\text{MoS}}_{2}$ monolayer},
  author = {Tahir, M. and Vasilopoulos, P. and Peeters, F. M.},
  journal = {Phys. Rev. B},
  volume = {93},
  issue = {3},
  pages = {035406},
  numpages = {9},
  year = {2016},
  month = {Jan},
  publisher = {American Physical Society},
  doi = {10.1103/PhysRevB.93.035406},
  url = {https://link.aps.org/doi/10.1103/PhysRevB.93.035406}
}

@article{Tutul2016,
  author  = {Tutul Biswas and Tarun Kanti Ghosh},
  title   = {Magnetotransport properties of the {$\alpha$-$T_{3}$} model},
  journal = {J. Phys.: Condens. Matter},
  volume  = {28},
  number  = {49},
  pages   = {495302},
  year    = {2016},
  doi     = {10.1088/0953-8984/28/49/495302}
}

@article{Aryan2026,
  title = {Magnetotransport properties of an unconventional Rashba spin-orbit coupled two-dimensional electronic system},
  author = {Pandita, Aryan and Islam, SK Firoz},
  journal = {Phys. Rev. B},
  volume = {113},
  issue = {11},
  pages = {115419},
  numpages = {11},
  year = {2026},
  month = {Mar},
  publisher = {American Physical Society},
  doi = {10.1103/3ytx-9fmt},
  url = {https://link.aps.org/doi/10.1103/3ytx-9fmt}
}

@article{Islam2018,
  author  = {S. K. Firoz Islam},
  title   = {Magnetotransport properties of 8-Pmmn borophene: effects of Hall field and strain},
  journal = {J. Phys.: Condens. Matter},
  volume  = {30},
  number  = {27},
  pages   = {275301},
  year    = {2018},
publisher = {IOP Publishing},
  doi     = {10.1088/1361-648X/aac8b3}
}

@article{GusyninPRL,
  title = {Unconventional Integer Quantum Hall Effect in Graphene},
  author = {Gusynin, V. P. and Sharapov, S. G.},
  journal = {Phys. Rev. Lett.},
  volume = {95},
  issue = {14},
  pages = {146801},
  numpages = {4},
  year = {2005},
  month = {Sep},
  publisher = {American Physical Society},
  doi = {10.1103/PhysRevLett.95.146801},
  url = {https://link.aps.org/doi/10.1103/PhysRevLett.95.146801}
}

@article{xzfz-d19q,
  title = {Integer quantum Hall effect in bilayer Dice lattices},
  author = {Liu, Han-Lin and Wang, J.},
  journal = {Phys. Rev. Res.},
  volume = {7},
  issue = {3},
  pages = {033195},
  numpages = {10},
  year = {2025},
  month = {Aug},
  publisher = {American Physical Society},
  doi = {10.1103/xzfz-d19q},
  url = {https://link.aps.org/doi/10.1103/xzfz-d19q}
}

@article{Streda1982,
  author  = {Streda, P.},
  title   = {Theory of quantised Hall conductivity in two dimensions},
  journal = {J. Phys. C: Solid State Phys.},
  volume  = {15},
  pages   = {L717},
  year    = {1982},
  doi     = {10.1088/0022-3719/15/22/005}
}

@article{PhysRevB.111.075125,
  title = {Quantum Hall effect and optical magnetoconductivity of two-dimensional topological nodal-line semimetals},
  author = {Barati, Shahin and Rahimpoor, Hamid and Abedinpour, Saeed H.},
  journal = {Phys. Rev. B},
  volume = {111},
  issue = {7},
  pages = {075125},
  numpages = {8},
  year = {2025},
  month = {Feb},
  publisher = {American Physical Society},
  doi = {10.1103/PhysRevB.111.075125},
  url = {https://link.aps.org/doi/10.1103/PhysRevB.111.075125}
}

@article{PhysRevLett.123.196403,
  title = {Berry Curvature Dipole in Strained Graphene: A Fermi Surface Warping Effect},
  author = {Battilomo, Raffaele and Scopigno, Niccol\'o and Ortix, Carmine},
  journal = {Phys. Rev. Lett.},
  volume = {123},
  issue = {19},
  pages = {196403},
  numpages = {5},
  year = {2019},
  month = {Nov},
  publisher = {American Physical Society},
  doi = {10.1103/PhysRevLett.123.196403},
  url = {https://link.aps.org/doi/10.1103/PhysRevLett.123.196403}
}

 \end{document}